# LASERS FOR COMMUNICATION AND COORDINATION CONTROL OF SPACECRAFT SWARMS

Himangshu Kalita,[*] Leonard Dean Vance,[†] Vishnu Reddy,[‡] and Jekan Thangavelautham[§]

Swarms of small spacecraft offer whole new capabilities in Earth observation, global positioning and communications compared to a large monolithic spacecraft. These small spacecrafts can provide bigger apertures that increase gain in communication antennas, increase area coverage or effective resolution of distributed cameras and enable persistent observation of ground or space targets. However, there remain important challenges in operating large number of spacecrafts at once. Current methods would require a large number of ground operators monitor and actively control these spacecrafts which poses challenges in terms of coordination and control which prevents the technology from scaled up in cost-effective manner. Technologies are required to enable one ground operator to manage tens if not hundreds of spacecrafts. We propose to utilize laser beams directed from the ground or from a command and control spacecraft to organize and manage a large swarm. Each satellite in the swarm will have a customized "smart skin" containing solar panels, power and control circuitry and an embedded secondary propulsion unit. A secondary propulsion unit may include electrospray propulsion, solar radiation pressure-based system, photonic laser thrusters and Lorentz force thrusters. Solar panels typically occupy the largest surface area on an earth orbiting satellite. A laser beam from another spacecraft or from the ground would interact with solar panels of the spacecraft swarm. The laser beam would be used to select a 'leader' amongst a group of spacecrafts, set parameters for formation-flight, including separation distance, local if-then rules and coordinated changes in attitude and position.

## INTRODUCTION

The rapid rise of small spacecraft and CubeSats in Low Earth Orbit (LEO) has increased accessibility, introducing new players to space exploration and enabling new commercial opportunities. At altitude below 450 km, the spacecraft face rapid decay in altitude due to aerodynamic drag and end up burning-up and disintegrating in the atmosphere within 1-2 years. With expected further advancement in electronics and increased congestion at lower altitudes, small spacecraft and CubeSats will begin to occupy higher altitudes in LEO. This is expected to include constellations of CubeSats to perform Earth observation, provide internet access, communications, Position, Navigation and Timing (PNT) and military services. New approaches are needed to dispose of and perform traffic management of these small satellites and CubeSats to prevent congestion, formation of debris fields and rise of the "Kessler Effect."

[*] PhD Student, Aerospace and Mechanical Engineering, University of Arizona, 85721, USA.
[†] PhD Student, Aerospace and Mechanical Engineering, University of Arizona, 85721, USA.
[‡] Associate Professor, Lunar and Planetary Laboratory, University of Arizona, 85721, USA.
[§§] Assistant Professor, Aerospace and Mechanical Engineering, University of Arizona, 85721, USA.

One commonly suggested strategy to moving or collecting of space debris is the use of specialized servicing/disposer spacecraft to perform rendezvous, capture and manipulation. However, this presents operational complexity and risks when interacting and making physical contact with some of these derelict spacecrafts that maybe damaged, spilling toxic propellants or containing spent radioactive waste.

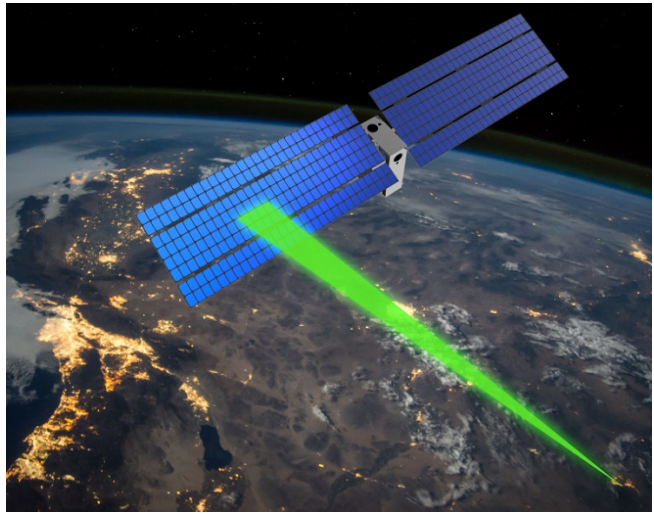

Figure 1: Lasers from the ground or space will interact with solar panels of the derelict spacecraft to power and it as part of space traffic management activities.

In this paper, we present an alternative approach to external servicing and space traffic management, where each spacecraft is plated with a "smart skin" containing solar panels, power and control circuitry together with an embedded secondary propulsion unit [17, 18]. Compared to previous papers, we go into depth in the application of this technology for swarm control, security in place to prevent hacking and testing of this technology on laboratory hardware. A secondary propulsion unit may include electrospray propulsion, solar radiation pressure-based system, photonic laser thrusters and Lorentz force thrusters. All of these propulsion systems either require minimal fuel or are propellant-less. Solar panels typically occupy the largest surface area on an earth-orbiting satellite. Furthermore, our previous work has shown that commercial space-grade solar panels can be used to detect and distinguish violet laser beams even when exposed to sunlight [2,16].

A laser beam from another spacecraft or from the ground would interact with solar panels of the derelict spacecraft. The "smart skin" would recognize gestural movements used to encode universal external positioning commands. The laser beam would be used to simultaneously communicate a 'move' and trigger operation of the secondary propulsion unit. The solar-panels in turn will power the smart-skin to permit these communication and command procedures. The laser beam maybe used to guide the movement of the spacecraft, trigger impulse maneuver commands, perform attitude control maneuvers and corrections. Ground and/or space surveillance would be used for verification, to start and stop movement, perform corrections and other such maneuvers. Use of laser beams to perform this external command and control offers some unique security benefits. The laser beams can be readily encrypted and because its directional and focused (i.e. from point to point), it is far less prone to eaves-dropping or hacking from a third-party.

This proposed approach facilitates staged intervention by a space traffic management organization to not only monitor, but also support providing commands to reposition satellites to prevent unwanted collisions or in the extreme case external commandeering of the derelict or damaged satellites to eliminate risks of collisions. This framework may also be applied for human command and control of satellite swarms that need to be maintain close formation while avoiding collisions. The use of human gestures enables intuitive interaction with these spacecrafts and should minimize fatigue and controller confusion after extended, strenuous intervention/commandeering. In the following sections we present background on the use of lasers for space communication, command and control, followed by presentation of the system architecture, description of the gesture control framework, use of laser ranging, external power transmission, discussions, conclusions and future work.

## BACKGROUND

Laser communication compared with traditional radio frequency communication methods provides much higher bandwidth with relatively small mass, volume and power requirements because laser enable the beams of photons to be coherent over large distances. LADEE demonstrated the advantages of laser communication, providing high bandwidth for a relatively small sized spacecraft [1]. However, LADEE utilized laser system onboard the spacecraft to perform high-speed bidirectional communication and consumes between 50 and 120 Watts. This is too high for spacecraft that typically produce a total power of less than 20 Watts.

Our previous work has shown a bi-directional communication system on a spacecraft without the need for a laser on the spacecraft itself [2, 16]. It has also shown that commercial space-grade solar panels can be used to detect and distinguish blue and violet laser beams even when exposed to sunlight. In our current approach, a laser beam will be used to directly communicate and control a derelict or inactive satellites and structures floating in orbit. With a customized "smart skin" containing solar panels, power and control circuitry and an embedded secondary propulsion unit onboard a spacecraft we can trigger a maneuver by sending a laser signal in the form of a gesture command from a ground station or another orbiting spacecraft.

Sending stroke gesture commands using a simple pointing device is common in various computer applications like marking menus with a pointing device [3]. Stroke gesture recognition is also used to send instructions to robots [4], develop robotic interface by free hand stroke [5]. Laser pointers has also been used extensively to send gesture commands to computers such as point-and-click or drag-and-drop [6,7]. It has also been used to tell a robot which object to pick up [8], which button to push [9] and also been used to specify target objects and give commands to robots to execute accordingly [10].

Satellite formation flying using environmental forces has also been studied extensively. Use of differential aerodynamic drag for satellite formation flying using drag plates has been studied by many researchers [11]. Similarly, satellite formation control using differential solar pressure with the help of solar flaps has also been studied [12]. Moreover, the use of geomagnetic Lorentz force as a primary means of spacecraft propulsion for satellite formation flying is also a well-studied area [13]. Techniques for detecting on-orbit satellites using laser ranging with centimeter accuracy has been shown [14]. These techniques will be used to identify the on-orbit derelict satellites and send maneuver control commands. Moreover, solar panels have also been used as a simultaneous wake-up receiver and for power harvesting using visible light communication [15].

## SYSTEM ARCHITECTURE

The proposed communication architecture consists of a customized "smart skin" containing solar panels, power and control circuitry and an embedded secondary propulsion system. A laser is beamed from a ground station or another spacecraft towards the satellite and the onboard photovoltaics acts as a wake-up laser receiver. This approach enables a laser ground station or a spacecraft to broadcast commands to the spacecraft in times of emergency that would trigger operation of the secondary propulsion system to perform impulse maneuvers, attitude control maneuvers and corrections. Moreover, adding an actuated reflector to the spacecraft will enable laser ranging and a two-way communication between ground station and the spacecraft, but without the laser diode being located on the spacecraft.

Fig. 2 shows the general systems architecture that is extended from [2, 16], between a ground station and an orbiting spacecraft. The key difference is that the array of receivers can detect spatial information, particularly what cell the laser beam has hit. The ground station is equipped with a microcontroller, a laser transmitter, an adaptive optics system, an array of laser receiver, a series of filters and a series of direction actuators. To mitigate the effect of atmospheric turbulence, the adaptive optics system, together with a reference laser beam is used to measure the beam's distortion when going through the atmosphere and compensate for the distortion by adjusting in the deformable mirror of the adaptive optics system. Direction actuators are used to point the laser transmitter and the receiver array towards the target spacecraft. The laser transmitter can send modulated laser beam to the target spacecraft. The receiver array receives the reflected laser beam and then filters it to gain maximum SNR using the micro-controller.

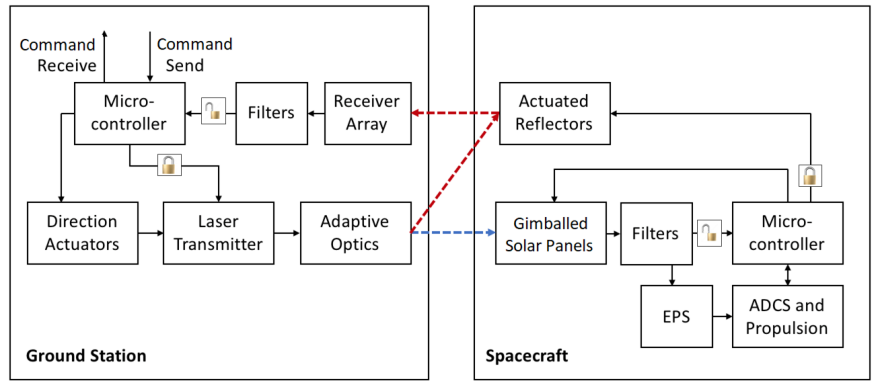

**Figure 1: Ground Station to Spacecraft System Bi-directional Architecture**

On board the spacecraft, the solar photovoltaic panels act as the laser beam receiver. The received signal is then processed through the filters and the DC component and the communication signal is separated using a bias tree. The DC component is transmitted to the onboard EPS system for power harvesting. The communication signal is processed through the microcontroller to gain maximum SNR and the telemetry data is processed to trigger the onboard ADCS and propulsion system. Fig. 3 shows the system architecture between two orbiting satellite.

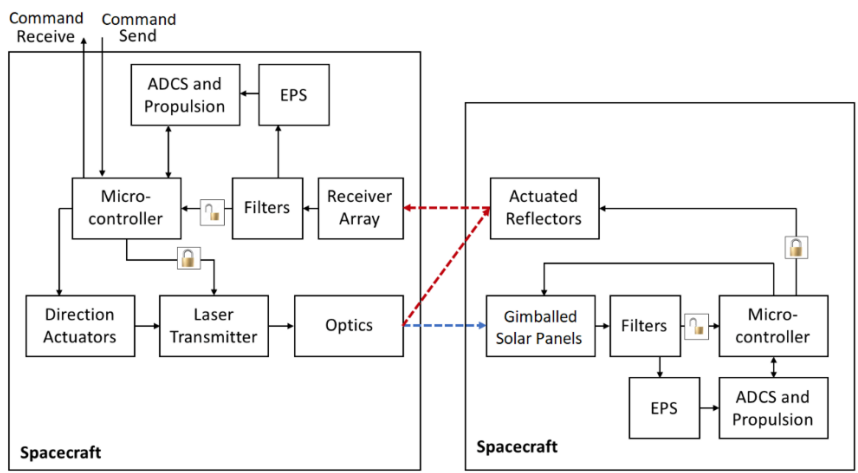

**Figure 2: Spacecraft to Spacecraft System Bi-directional Architecture**

Spacecraft 1 is equipped with a microcontroller, a laser transmitter, an adaptive optics system along with a series of direction actuators to send a gesture command through a laser signal while spacecraft 2 is equipped with a microcontroller and gimballed solar to identify the gesture command and trigger a maneuver. Moreover, an encryption layer is added for data and commands just before being sent to the laser transmitter. Decryption is performed after the signal is filtered and ready to be interpreted by the micro-controller. Through this encryption/decryption process access to the spacecraft is only possible thanks to the right set of passcodes shared between ground control and spacecraft. The passcode for encryption and decryption maybe one or a few gestures prompted at the beginning of a message/command or passed through as a modulatory signal. The passcode would then be used to decrypt the message and perform verification. When verification fails, the commanded message/communication is ignored, or systems goes into safe-mode after too many wrong tries. Fig. 4 and Fig. 5 show uni-directional laser communication between ground and spacecraft and spacecraft and spacecraft respectively.

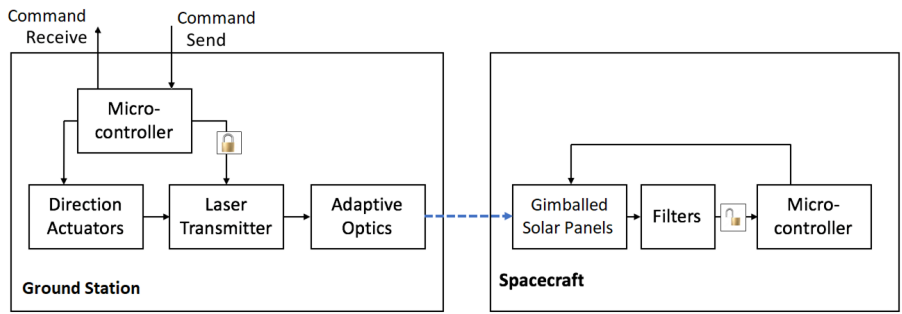

Figure 3: Ground Station to Spacecraft Uni-directional Architecture.

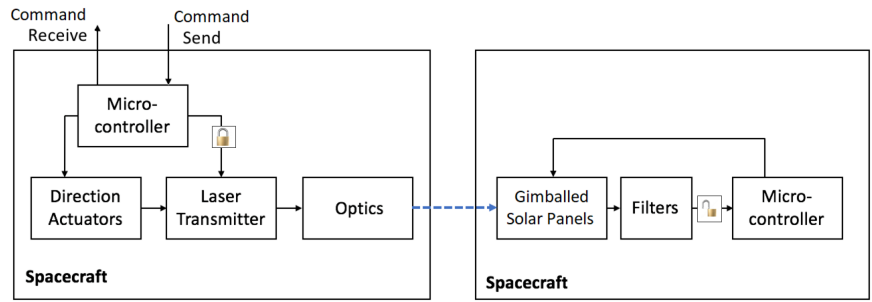

Figure 4: Spacecraft to Spacecraft Uni-directional Architecture.

## GESTURE CONTROL

Gestures are increasingly becoming a predominant mode of human-machine interaction. This is principally due to them being intuitive, requiring minimal training. Stroke gestures also sometimes called "pen gestures" represents the movement trajectory of one or more contact points on a sensitive surface. The most significant advantage of using stroke gestures to input commands in that the user can specify several kinds of commands using just a simple pointing device. In our case, a laser beam would be used as a pointing device with the "smart-skin" acting as the sensitive sensing surface. A laser beam from another spacecraft would interact with the solar panels of the derelict spacecraft.

The laser beam would be used to communicate a 'move' which would then trigger operations on the derelict spacecraft. The laser beam maybe used to guide the movement of the spacecraft, trigger impulse maneuver commands, perform attitude control maneuvers and corrections. This method of gesture control will be used to control a cluster of closely flying satellite and execute satellite formation flying. One of the most important challenges of the satellite formation flying involves controlling the relative positions of the satellites in the presence of external disturbances, i.e., gravitational perturbation including the Earth's oblateness ($J_2$ effect), aerodynamic drag, and solar radiation pressure.

These issues can be addressed by the use of environmental forces including differential aerodynamic drag, differential solar radiation pressure, and Lorentz force. The satellite formation flying system comprises of a leader and follower satellites equipped with either drag plates, solar flaps or Lorentz actuation system. The orbital equations of motion for the leader satellite and the relative equations of motion of the follower satellites are as follows:

$$\ddot{r}_c = r_c \dot{\theta}^2 - \frac{\mu}{r_c^2}, \qquad \ddot{\theta} = -\frac{2\dot{\theta}\dot{r}_c}{r_c} \tag{1}$$

$$m_f \ddot{x} - 2 m_f \dot{\theta} \dot{y} - m_f (\dot{\theta}^2 x + \ddot{\theta} y) + m_f \mu \left\{ \frac{(r_c + x)}{r^3} - \frac{1}{r_c^2} \right\} = f_x + f_{dx} \tag{2}$$

$$m_f \ddot{y} + 2 m_f \dot{\theta} \dot{y} + m_f (-\dot{\theta}^2 y + \ddot{\theta} x) + m_f \frac{\mu}{r^3} y = f_y + f_{dy} \tag{3}$$

$$\ddot{z} = -\frac{\mu z}{r^3} + f_z + f_{dz} \tag{4}$$

The leader satellite is in a reference orbit that is assumed to be planar and defined by a radial distance $r_c$ from the center of the Earth and a true anomaly $\theta$. The follower satellite moves in a relative trajectory about the leader satellite, in a relative frame $xyz$ fixed at the leader satellite as shown in Fig. 6. In the (2), (3) and (4) $m_f$ denotes the mass of the follower satellite, $f_{dx}$, $f_{dy}$, and $f_{dz}$ are the disturbance forces and $f_x$, $f_y$ and $f_z$ are the control forces.

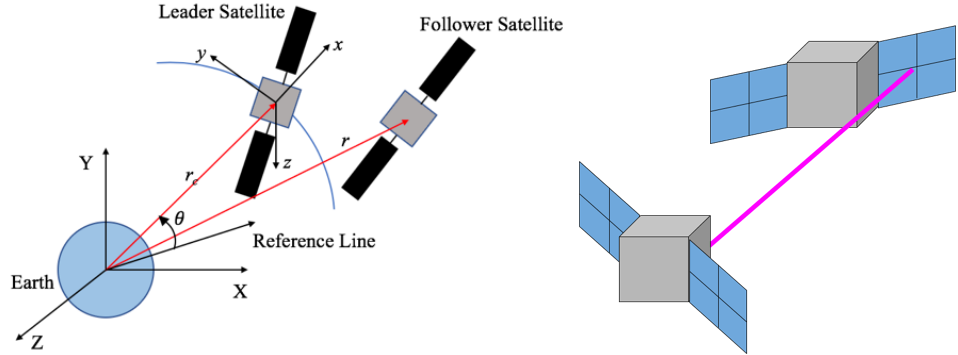

**Figure 5: (Left) Leader and Follower satellite reference frames. (Right) Leader satellite sending a gesture command to a follower satellite using laser beams.**

Three different types of desired formation trajectories are considered for this paper.

**Along Track Formation Flying (AF).** The follower shares the same ground track as the leader satellite. It has to keep a constant desired along track separation of $r_d$ and the desired trajectory is defined as:

$$y_d = r_d \tag{5}$$

**Projected Circular Formation Flying (PCF).** The leader and the follower satellite maintain a fixed relative distance only on the $yz$ plane and the formation is defined as $y^2 + z^2 = r_d^2$. The desired trajectory is defined as:

$$\begin{Bmatrix} x_d \\ y_d \\ z_d \end{Bmatrix} = \left(\frac{r_d}{2}\right) \begin{bmatrix} \sin(\dot{\theta}_m t + \varphi) \\ 2\cos(\dot{\theta}_m t + \varphi) \\ 2\sin(\dot{\theta}_m t + \varphi) \end{bmatrix} \quad (6)$$

**Circular Formation Flying (CF).** The leader and the follower satellite maintain a constant separation from each other, and the formation is defined as $x^2 + y^2 + z^2 = r_d^2$. The desired trajectory is defined as:

$$\begin{Bmatrix} x_d \\ y_d \\ z_d \end{Bmatrix} = \left(\frac{r_d}{2}\right) \begin{bmatrix} \sin(\dot{\theta}_m t + \varphi) \\ 2\cos(\dot{\theta}_m t + \varphi) \\ \sqrt{3}\sin(\dot{\theta}_m t + \varphi) \end{bmatrix} \quad (7)$$

Where $\varphi$ is the in-plane phase angle between the leader and the follower satellites, and $\dot{\theta}_m = \sqrt{\mu/a_c^3}$ is the mean angular velocity. We have identified command methods as single-stroke gestures for performing different satellite formation maneuvers. Fig. 7 shows stroke gestures representing along track formation flying (AF), projected circular formation flying (PCF), and circular formation flying (CF). The laser pointer on the leader satellite is mounted on a head that can move with fine precision using a SMA or piezoelectric actuation mechanism. The "smart-skin" can identify the laser hitting individual solar cells and hence identify the gesture stroke.

When the leader satellite draws a straight line along the solar panels, the along track formation flying (AF) maneuver is triggered, a clockwise circle triggers the projected formation flying (PCF) maneuver while a clockwise circle with a line along one of its diagonal triggers the circular formation flying (CF) maneuver. In addition to that, gesture strokes to cancel, undo and redo a maneuver is also identified as shown in Fig. 8. The lower row of gestures could be used by the spacecraft to record a sequence of gestures strokes into a macro. This includes the record macro, play macro and stop macro recording command.

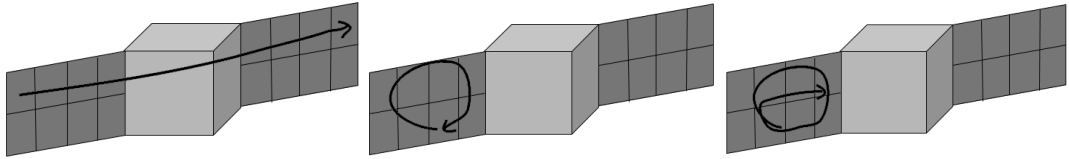

**Figure 6:** Gesture command strokes for, a) Along track formation flying (AF), b) Projected Circular Formation Flying (PCF), c) Circular Formation Flying (CF).

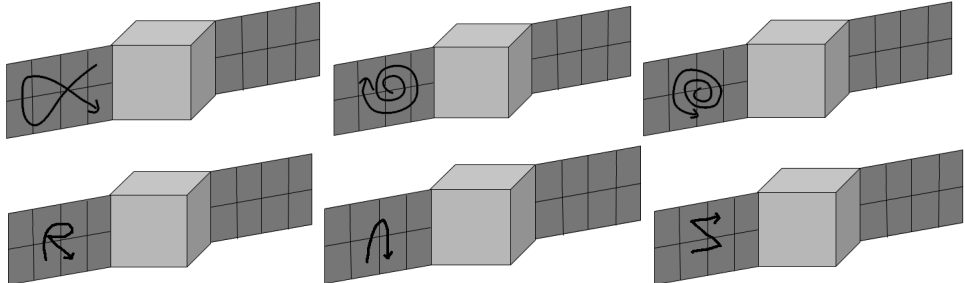

**Figure 7:** Gesture command strokes (upward row from left to right) to a) Cancel, b) Redo and c) Undo a maneuver and (bottom row from left to right) d) Record Macro e) Play Macro f) Stop Macro.

Ground and/or space surveillance would be used for verification, to start and stop movement, perform corrections and other such maneuvers. The entire move maneuver would be made possible without operation of the Command and Data Handling Computer onboard the derelict satellite. Thus, the laser beam would act as a 'remote control' for the spacecraft.

**Formation Flying.** For formation flight, a leader spacecraft is selected using gestures (Fig.9 top left) and this is followed by identification of the remaining spacecraft in the group Fig. 9 (top center) followed by locking the relative position of each spacecraft Fig. 9 (top right). After the group of spacecrafts are locked in relative position and attitude, then gestures movements applied to the leader spacecraft will result in the remainder of the spacecraft following the leader in tandem, maintaining fixed distance and attitude. Finally, Fig. 9 (bottom) shows a gesture to unlock a spacecraft in terms of relative position and attitude from the group.

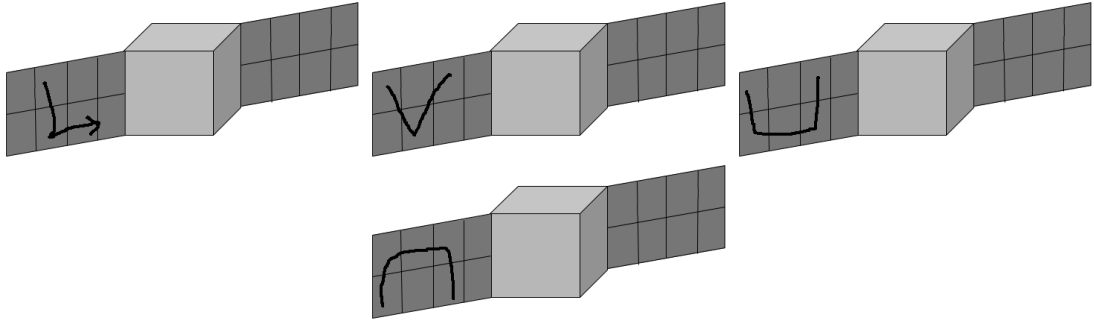

**Figure 8:** (Top Left) Gesture command strokes to select leader amongst a flock of spacecraft. (Top Center) Gesture to identify other spacecraft that are part of the current group. (Top Right) Gesture to lock relative position and attitude of each current group member spacecraft to the leader. (Bottom) Gesture to unlock relative position/attitude of a spacecraft from a group.

**Alphabet of Gestures.** Using this general approach, an alphabet of gestures representing symbols and high-level commands can be represented. The limits on the number of gestures is dependent on the solar-cell packing density (analogous to pixel density on a flat panel display) and signal processing frequency (to recognize speed of gesture movement). A third factor can be modulation of the signal.

**Modulation.** The laser beam maybe used to encode a signal through modulation [2, 16]. This modulation maybe used to encode for "intensity" without having to allocate a symbol in the alphabet. Applied with the gesture shown in Fig. 6., the intensity maybe proportional to the linear or angular velocity of the spacecraft. Applied with the play macro gesture, this may determine the replay speed.

## HARDWARE DEMONSTRATION

A simple laboratory testbed to demonstrate solar cell gesture control was built to explore the basic viability of this concept (Fig . 10). This vehicle mounts four small solar panels on an Arduino based robotic vehicle. The four solar panels provide an analog voltage signal to the Arduino controller which can monitor them for differential signals which would correspond to gestures provided through a laser pointer. Figure 9 shows the overall block diagram for this vehicle, and the following figure shows the layout of the actual vehicle.

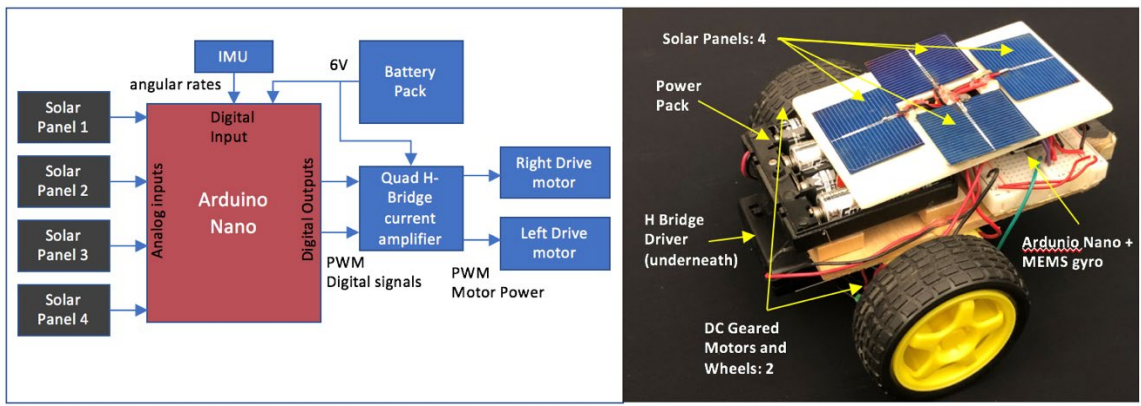

**Figure 9: Layout of demonstration vehicle.**

The testbed looks for sequential signal pulses between the fore and aft mounted solar panels to control speed, and the left and right panels to control heading (Fig. 11). When sequential pulses (from a laser pointer) are received in succession by the back and then the front panel, throttle is increased by 20%, up to the maximum of 100%. Likewise, when the left panel senses a pulse, followed by the right panel, the robot adjusts heading to the right by 30 degrees and holds it using feedback from the MEMS IMU package integrated with the Arduino. One other gesture is recognized which is the back panel receiving a long pulse, this is interpreted as a full-stop signal.

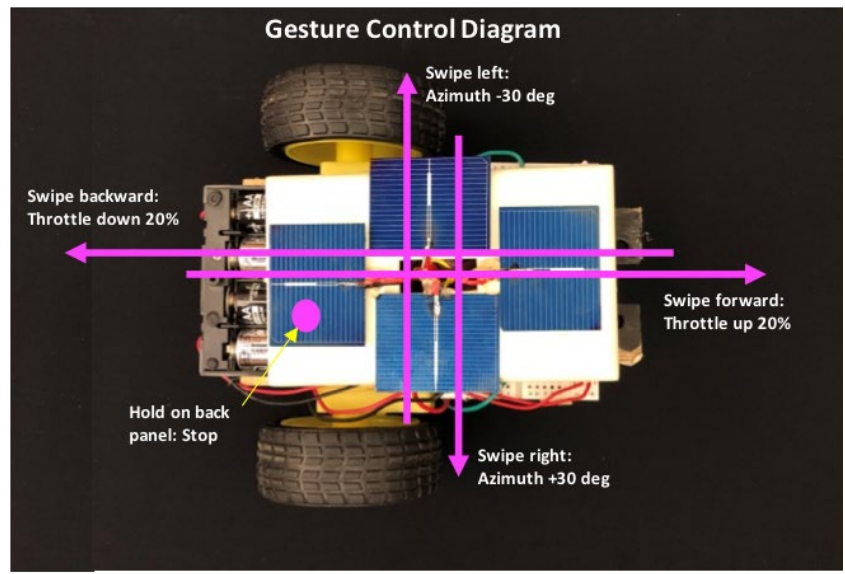

**Figure 10: Vehicle gesture control diagram.**

Using gestures from a laser pointer, an operator can then swipe the light beam forward, backwards, left and right to control the speed and direction of the robot. Complicated obstacles can be navigated easily, all using the voltage output directly from solar panels. The basic technique of sensing light pulses moving between panels can be extrapolated to more complex motions, permitting the creation of an ersatz language for communications with the robot vehicle.

**CONCLUSION**

In this paper, we presented a new systems architecture for external position control and traffic management of on-orbit derelict satellite by using a laser beam. In our approach, a laser beam will be used to directly communicate and control a derelict or inactive satellites and structures floating in orbit. The same approach maybe also used to actively command and control one or more satellites in a swarm. The satellite will have a customized "smart skin" containing solar panels, power and control circuitry and an embedded secondary propulsion unit. A laser beam from another spacecraft or from the ground would interact with solar panels of the derelict spacecraft in the form of gesture commands. The on-orbit satellite will recognize the gesture command and then would trigger operation of the secondary propulsion unit. The laser beam maybe used to guide the movement of the spacecraft, trigger impulse maneuver commands, perform attitude control maneuvers and corrections.

We have identified simple gesture commands to trigger along track formation flying, projected circular formation flying and circular formation flying maneuvers. Moreover, gesture commands to cancel, redo and undo a particular maneuver are also identified that would allow the laser beam to act as a remote control for the spacecraft. Laser ranging would be used for ground surveillance of these satellites that would allow us to start, stop or verify a maneuver. In case of a need for emergency power, power can be transmitted from the ground or from space by shooting a laser beam and the "smart-skin" operating as the power harvesting module.

The laser beam will enable a secure point to point communication and cannot be eavesdropped, unless if the eavesdropping unit is in the way or close to the derelict satellite. However, if RF (Radio Frequency) were to be used, then eavesdropping maybe possible without detection. RF signal requires licensing and is congested due to high demand. Use of a laser beam avoids these logistical challenges. This laser system may serve as a secure backup system that can be used to mitigate and take back control of a satellite from cybersecurity threats/hacking using RF communication. However, significant advancements are required to make this approach practical and efforts are underway to develop laboratory prototypes to validate the feasibility of our proposed systems architecture.

**REFERENCES**

1. D. M. Boroson, J. J. Scozzafava, D. V. Murphy, B. S. Robinson, "The Lunar Laser Communications Demonstration (LLCD)," 3$^{rd}$ IEEE International Conference on Space Mission Challenges for Information Technology, 2009.
2. X. Guo, J. Thangavelautham, "Novel Use of Photovoltaics for Backup Spacecraft Laser Communication System," IEEE Aerospace Conference, 2017.
3. G. Kurtenbach, W. Buxton, "User Learning and Performance with Marking Menus," SIGCHI Conference on Human Factors in Computing Systems, 1994.
4. D. Sakamoto, K. Honda, M. Inami, T. Igarashi, "Sketch and Run: A Stroke-based Interface for Home Robots," 27$^{th}$ International Conference on Human Factors in Computing Systems, 2009.
5. M. Skubic, D. Anderson, S. Blisard, D. Perzanowski, A. Schultz, "Using a hand-drawn sketch to control a team of robots," Autonomous Robots, 2007.
6. D. R. Olsen, T. Nielsen, "Laser pointer interaction," SIGCHI Conference on Human Factors in Computing Systems, 2001.
7. C. Kirstein, H. Muller, "Interaction with a Projection Screen using a Camera-Tracked Laser Pointer," International Conference on MultiMedia Modelling, 1998.

8   C. C. Kemp, C. D. Anderson, H. Nguyen, A. J. Trevor, Z. Xu, "A Point-and-Click Interface for the Real World: Laser Designation of Objects for Mobile Manipulation," 3$^{rd}$ ACM/IEEE International Conference on Human-Robot Interaction, 2008.
9   T. Suzuki, A. Ohya, S. Yuta, "Operation Direction to a Mobile Robot by Projection Lights," IEEE Workshop on Advanced Robotics and its Social Impacts, 2005.
10  K. Ishii, S. Zhao, M. Inami, T. Igarashi, M. Imai, "Designing Laser Gesture Interface for Robot Control," IFIP Conference on Human-Computer Interaction, 2009.
11  S. Varma, K. D. Kumar, "Multiple Satellite Formation Flying Using Differential Aerodynamic Drag," Journal of Spacecraft and Rockets, 2012.
12  S. Varma, K. D. Kumar, "Multiple Satellite Formation Flying using Differential Solar Radiation Pressure," AIAA/AAS Astrodynamics Specialist Conference, 2010.
13  S. Tsujii, M. Bando, H. Yamakawa, "Spacecraft Formation Flying Dynamics and Control Using the Geomagnetic Lorentz Force," Journal of Guidance, Control, and Dynamics, 2013.
14  J. J. Degnan, "Millimeter Accuracy Satellite Laser Ranging: A Review," Contributions of Space Geodesy to Geodynamics: Technology, 1993.
15  C. Carrascal, I. Demirkol, J. Paradells, "A novel wake-up communication system using solar panel and visible light communication," IEEE Global Communications Conference, 2014.
16  X. Guo, J. Thangavelautham, "Low-cost Long Distance High Bandwidth Laser Communication System for Small Mobile Devices and Spacecraft," Patent: US 9,991,957, 6 Jun, 2018.
17  H. Kalita, L. Vance, V. Reddy, J. Thangavelautham, "Laser Beam for External Position Control and Traffic Management of On-Orbit Satellites," Advanced Maui Optical and Space Surveillance Technologies Conference, 2018.
18  H. Kalita, L. Vance, J. Thangavelautham, "Laser Beam for External Position Control and Traffic Management of On-Orbit Satellites," Provisional Patent, US 62/731,399, 2018.